\documentclass[epj, twocolumn]{webofc}
\usepackage[varg]{txfonts}
\usepackage{enumitem}
\usepackage{graphicx}
\usepackage{epstopdf}
\usepackage[T1]{fontenc}
\usepackage{microtype}

\woctitle{ISVHECRI 2014}

\begin{document}
\title{The meaning of the UHECR Hot Spots}
\subtitle{A Light Nuclei Nearby Astronomy}

\author{Daniele Fargion\inst{1,2}\fnsep\thanks{\email{daniele.fargion@roma1.infn.it}},
        Graziano Ucci\inst{1} \and Pietro Oliva\inst{3} \and Pier Giorgio De Sanctis Lucentini\inst{4}}

\institute{Physics Depart., Rome University 1, Sapienza, Ple. A. Moro 2, 00185, Rome, Italy \and INFN, Ple. A. Moro 2, 00185, Rome, Italy \and Engin. Depart., Niccol\`o Cusano University, Via Don Carlo Gnocchi 3, 00166 Rome, Italy \and GUBKIN Uni. of Oil and Gas, Leninsky prospekt 65, Moscow, Russia.}

\abstract{In this paper we review all the up-to-date Ultra High Energy Cosmic Ray (UHECR) events reported by AUGER, the  Telescope Array (TA) and AGASA in common coordinate maps.
We also confirm our earliest (2008-2013) model, where UHECRs  mostly comprise light nuclei (namely He, Be, B), which explains the Virgo absence and confirms M82 as the main source for the Northern TA Hot Spot. Many more sources, such as NGC253 and several galactic sources, are possible candidates for most of the 376 UHECR events. Several correlated maps, already considered in recent years, are  reported to show all the events,  with their statistical correlation values.}

\maketitle

\section{Introduction}
In November 2007, AUGER \cite{Auger-Nov07} reported, within 27 events, a rare clustering of UHECR events centered around Cen-A with a spread of $\pm 15^\circ$: a first Southern Hot Spot anisotropy.
The later 2011 data (69 UHECR events, as well as the additional train of dozens of twin events at 20 EeV \cite{Auger-Nov11} \cite{Fargion2011nima}) confirmed  such a UHECR Hot Spot, thus strengthening the nuclei-like (not the proton-like) UHECR composition supposition \cite{Fargion2008}.
Since 2011, neither the Virgo Cluster nor any Super-Galactic anisotropy (as had been first considered to be detectable assuming UHECR proton composition) were in fact detected by AUGER. On the other hand the Telescope Array (TA)   found, in recent  months, a comparable remarkable anisotropy (a Northern Hot Spot, also spread over $\pm15^\circ$) pointing to an uncorrelated and unexpected region of the sky. Therefore, up to now we only took  care  of two Hot Spots in the Northern and Southern Sky.
Moreover and once again unnoticed, the Virgo cluster - better viewed from TA skies - did not manifest itself at all, even in the TA data, where one would expect it to show if, as TA claims, UHECRs are protons.
To make  this race in discovering UHECR Astronomy more contradictory, we must remember that since 2007 AUGER favored an iron composition, while TA always supported protons. Eventually, the most recent TA  events ($72+15$ events) and the deeper and final AUGER records (231 events in the 2004-2014 period) \cite{Auger-Nov14} offer a unique opportunity to combine their maps and test eventual sky correlations though AUGER disclaimed any possible relevant connection with the Cen-A Hot Spot.

The AUGER and TA maps are often in non-comparable map projections. Here we  combined both of those published (for TA in-the-press) events in a known and useful coordinate system. We shall discuss and weight the clustering along these maps overlapping on different ones.
It should be remembered that, since very early 2008 \cite{Fargion2008} (see also \cite{Fargion2011nima}),  we proposed that a possible solution to the AUGER puzzle (anisotropy size and Virgo absence) could be if UHECRs were mostly the lightest nuclei, for example helium.
Such a composition would indeed explain the spread angle of UHECR by their random bending  along galactic fields ruling out a unique Hot Spot due to our nearest AGN: Cen-A. Moreover,  He-like nuclei are fragile through photo-nuclear dissociation by the Cosmic Microwave Background: they cannot cross large GZK distances as can protons  ($50-80$ Mpc), but they can fill a few smaller Universe size ($3-4$ Mpc). Because of this Cen-A could shine, Virgo could not.
The lightest UHECR nuclei may also explain, as we argued then and now, the remarkable earliest AUGER and recent TA results, the Virgo absence, by the lightest fragility and opacity: in the same model the TA Hot Spot in the Northern Sky must  originate from a very nearby source. The Ursa Major Cluster or Virgo are too far, but a closer source such as the M82 star-burst galaxy could survive the distance.
\begin{figure}[t]
\centering
\includegraphics[width=0.48\textwidth,clip]{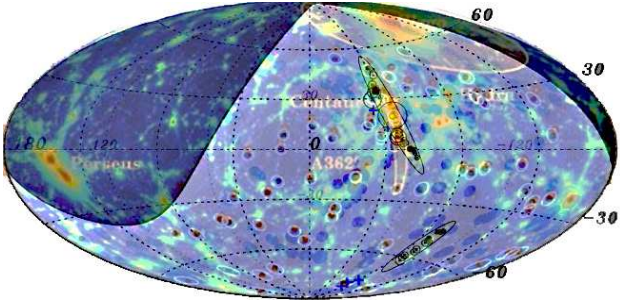}
\caption{An early (2011) UHECR AUGER data above GZK, cut by 69 Auger events and 20 EeV, clustering on Infrared background: note the Virgo absence and the Cen-A clustering \cite{Fargion2010}.}
\label{fig-1}
\end{figure}
\begin{figure}[t]
\centering
\includegraphics[width=0.48\textwidth,clip]{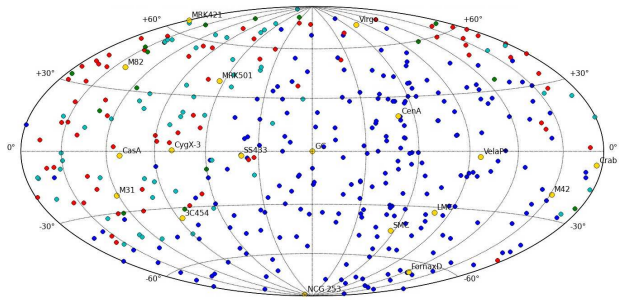}
\caption{Hammer Projection in Galactic coordinates for the latest 231 AUGER UHECR (blue) with additional 72+15=87 TA (72 blue and 15 green) and 58 AGASA (cyan) records, in a total of 376 events. A  few potential sources are labeled.}
\label{fig-2}
\end{figure}

The so-called TA Hot Spot (about a dozen  events) is not centered on M82, but is coherently bent by ($15-20^\circ$) in nearby Northern Galactic fields (see fig.\ref{fig-2}-\ref{fig-5} and fig.\ref{fig-magnetic}-\ref{fig-magnetic2}). However, we recognize in the same bending tail a much nearer, narrower and very recent (this year) a 5 event clustering, confirming such a probable coherent bending of UHECR light nuclei from M82. Such a quintuplet cluster within 100 square degrees has a chance probability of  $<10^{-3}$ to occur. Moreover the oldest UHECR AGASA record in the  Northern sky (58 events)  showed a unique triplet almost overlapping, making the chance probability to find such a clustering for 8 events within 150-200 square degrees, as low as $(1\div4)\cdot10^{-4}$. We will reconfirm  that Cen-A, M82 and a few galactic sources like Vela, Cygnus X3 and SS433, could eject light or heavier UHECR nuclei, possibly radioactive ones, bent by large local magnetic fields. We were inspired by  early preliminary UHECR correlations with TeV anisotropy maps discovered in the last decade by Milagro, ARGO and now more recently  by HAWK and ICECUBE. We  again claim that such TeV-UHECR correlations might be due to  UHECR fragment nuclei by radioactive decay in flight and/or by UHECR lightest nuclei secondaries photo-dissociation in their travels. We note a new interesting clustering along the South Galactic Pole pointing to the main nearby star-burst galaxy NGC 253, an object  similar to  M82, once again within 3 Mpc distance.

\subsection{Cen-A, M82 and NGC 253 as the main extragalactic UHECR sources}

\begin{figure}[h]
\centering
\includegraphics[width=0.48\textwidth,clip]{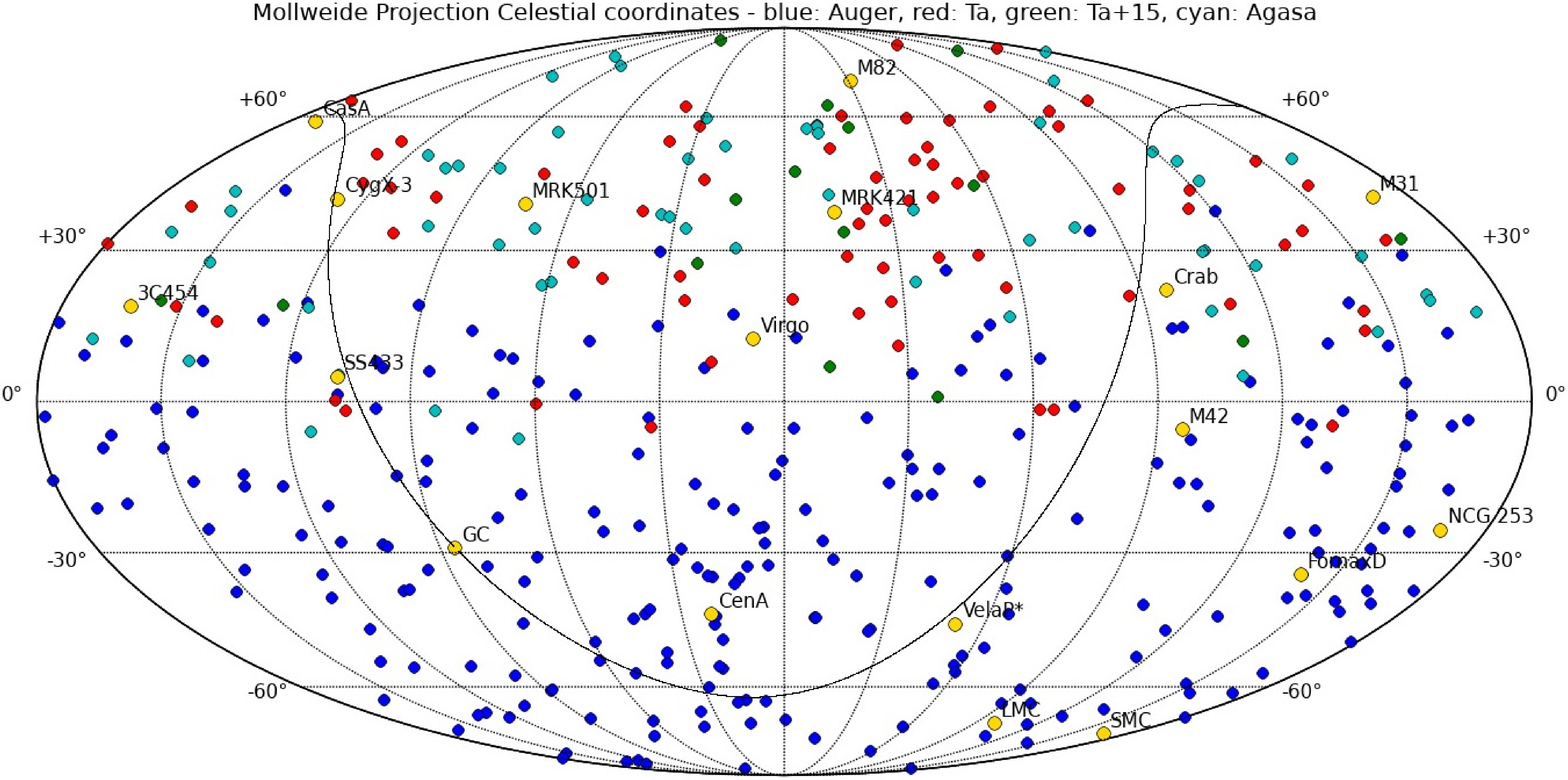}
\caption{Mollweide Projection in Celestial coordinates for the latest 231 AUGER UHECR (blue) with additional 72+15=87 latest TA (72 red and 15 green) records. A few potential sources are labeled.}
\label{fig-3}
\end{figure}

\begin{figure}[t]
\centering
\includegraphics[width=0.48\textwidth,clip]{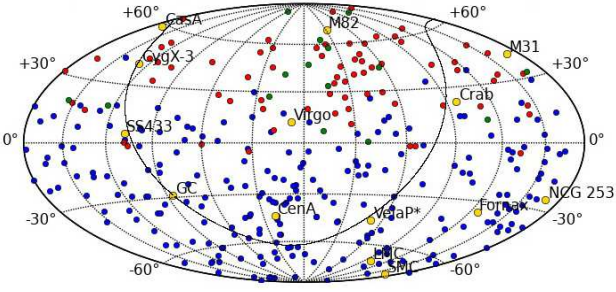}
\caption{Hammer Projection in Celestial coordinate for the latest 231 AUGER UHECR (blue) with additional 72+15=87 latest  TA (72 red and 15 green) records. A few potential sources are labeled as well as the galactic plane.}
\label{fig-4}
\end{figure}

\begin{figure}[t]
\centering
\includegraphics[width=0.48\textwidth,clip]{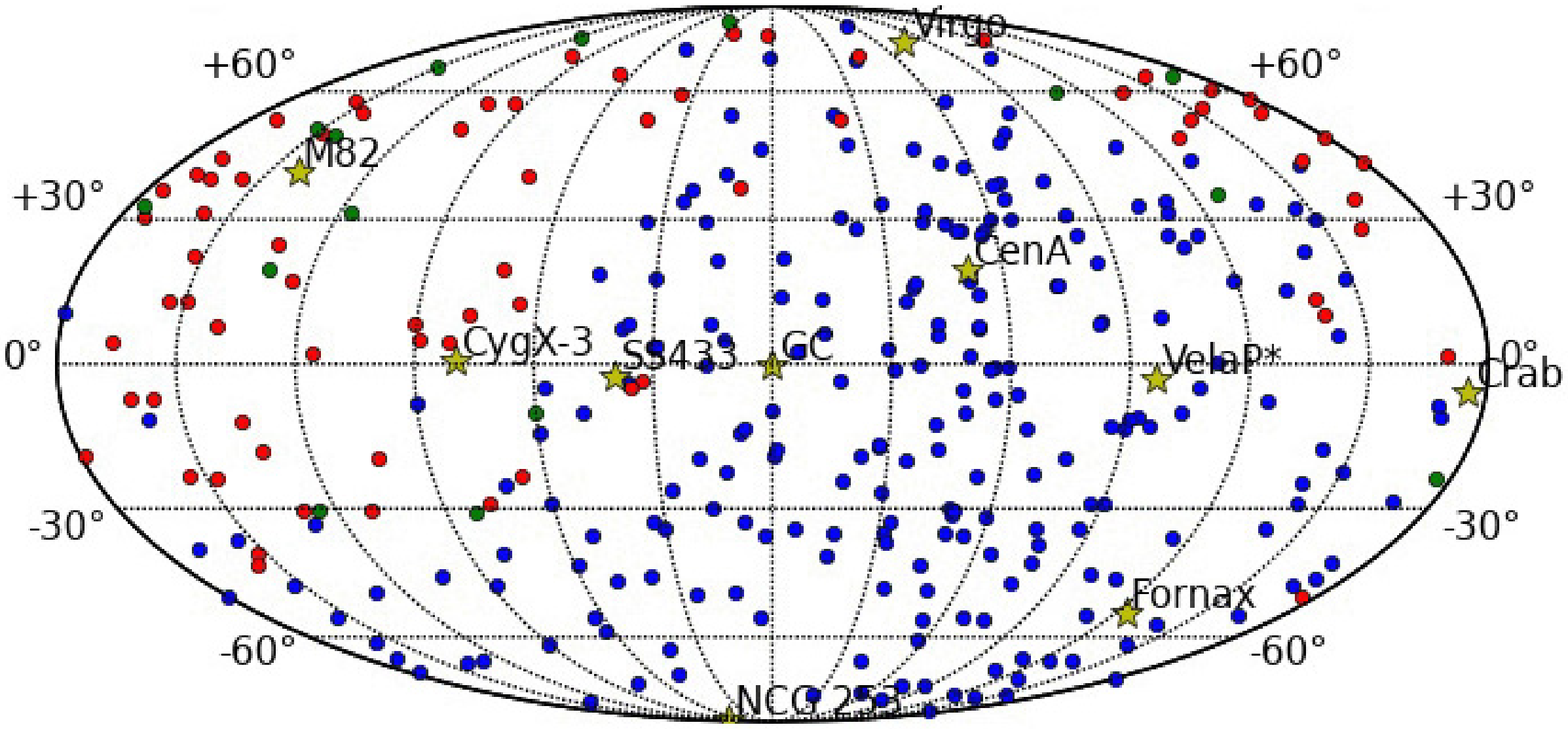}
\caption{Mollweide Projection in Galactic coordinates for the latest 231 AUGER UHECR (blue) with additional 72+15=87 latest TA (72 red and 15 green) records. A few potential sources are labeled.}
\label{fig-5}
\end{figure}

UHECRs are clustering both within the Southern (AUGER) and  Northern (TA) skies in wide $\pm 15^\circ$ Hot Spots. The nearest Galactic Cluster, Virgo, within the expected GZK opacity distance for protons (related to the apparently observed GZK cut off observed spectra by Hires, AUGER and TA) is absent. This absence is remarkable in particular in recent TA data as well as in the latest 231 AUGER events. The Virgo clustering absence does not fit with UHECRs as nucleons. The same $\pm 15^\circ$ Hot Spot clustering cannot fit with an extra-galactic heavy nucleus composition (Fe, Ni) because of their much larger charge and wider deflection angle (above $90^\circ$). However, the lightest UHECR nuclei, namely or mostly He, are fragile enough to be absorbed and soon hidden  by photo-dissociation on cosmic radiation on their 20 Mpc journey from Virgo, thus explaining their absence. On the contrary, AUGER clustering around the nearest active AGN, Cen-A (3.5 Mpc), may be caused by  random \emph{incoherent} bending of light He nuclei along the galactic magnetic fields; the absence of Virgo in the TA data is explained again by the UHECR He opacity.
The TA Hot Spot clustering could be ejected by the star-burst M82 (3.5 Mpc) as the main source, while UHECRs are bent and spread by a \emph{coherent} magnetic field either galactic at the North Pole and/or extra-galactic. We want to emphasize here the clustering of 5 UHECR TA events nearer to M82, since it might be a first UHECR signal around this main source. This very narrow spot (5 events)  almost overlaps an older one: a triplet observed by AGASA in 1990-2000. The binomial probability to find 4+1  such events inside a narrow area of $10^{2}$  square degrees, amongst 87 TA signals, is $< 8.2\cdot10^{-4}$, even ignoring the AGASA triplet.
Also considering  the additional $58$ AGASA events and a triplet clustering within $150^\circ$, the probability shrinks to $10^{-4}$. These quintuplet signals are additional to the remaining (more deflected) He-like nuclei found in the wider TA hot spot (21 events among 87 in nearly $2000$ square degree sky), whose chance probability to occur is as low as $< 2 \cdot 10^{-4}$ but their clustering is more diluted and far from M82. Additional signals might be gamma secondaries of these UHECR He photo-dissociation in flight (or radioactive decay in flight as for the He$^{6}$ isotope or Be$^{7}$ or the more abundant and deflected Al$^{26}$) that might also point and trace, by a boosted Lorentz factor, part of these gamma anisotropies on TeV  maps discovered by Milagro, ARGO, HAWK and ICECUBE. A few UHECR sources might also be  galactic as the very recent multiplet (8 events) along Vela also shows  Cygnus X3 clustering raised as a peculiar narrow multiplet as well as a much narrower clustering event (around SS433 or Aq1) and possible spread sources along the Magellanic stream regions (LMC and SMC, or NGC 253, a second near - 3.5 Mpc - star-burst source and/or Fornax Dwarf Galaxy source),
see Figures \ref{fig-2}-\ref{fig-5}, \ref{fig-13} and \ref{fig-magnetic}-\ref{fig-magnetic2}.

\subsection{Bending for He UHECR and fragments at 20 EeV along Cen-A}

Let us remember why we should choose a He-like composition: the old 2011 UHECR multiplet clustering foreseen in 2009  \cite{Fargion2011nima} and observed in 2011 \cite{Auger-Nov11} (and forgotten in most reports) by AUGER UHECR at 20 EeV energy contains just three apparently isolated trains of events which point to unknown sources (see old 69 events on IR map, see fig.\ref{fig-1}).
However, the crowding of the two train multiplet tail fixed inside a very narrow disk area focused about the Cen-A UHECR source  is  remarkable as it had already been foreseen  \cite{Fargion2011c}.
If UHECRs are composed of protons (as some AUGER and TA authors believe), they will not naturally explain such a tail structure because these events do not cluster more than a few degrees, unlike the observed UHECRs and the associated multiplet. He-like UHECRs fit the AUGER composition traces as well as the HIRES and TA ones. The He secondaries split in half (or in fourths) energy fragments along the Cen-A tail, the presence of was  been recently foreseen  \cite{Fargion2011c}.
Indeed, the dotted circle around Cen-A containing two (of three) multiplets in a radius as small as $7.5^\circ$ extends in an area that is as small as $200$ square degrees, below or near 1\% of the AUGER observation sky. The probability that two  of these  three sources fall inside this foreseen small area is by the binomial distribution $\simeq 3 \cdot 10^{-4}$. Moreover the same twin tail of the events is aligned almost exactly $\pm0.1$ rad along the UHECR train of events toward Cen-A. Therefore the UHECR  multiplet alignment  at 20 EeV has a probability as low as $P (3,2) \simeq 3 \cdot 10^{-5}$ of following an
\emph{a priori} foreseen signature (\cite{Fargion2011nima},   \cite{Fargion2011c} and  \cite{Fargion2010}).

The incoherent random angle bending along the galactic plane and arms, $\delta_{rm}$, while crossing along the whole Galactic disk ($L \simeq 20$ kpc)  in different, alternating, spiral arm fields  and within a characteristic  coherent length  $l_c \simeq2$ kpc for He nuclei is
$$
\delta_{rm-He}\simeq16^\circ\frac{Z}{Z_{\textrm{He}^2}}\cdot\left(\frac{6\cdot10^{19}\,\textrm{eV}}{E_{CR}}\right)\left(\frac{B}{3\,\mu \textrm{G}}\right)\sqrt{\frac{L}{20\, \textrm{kpc}}} \sqrt{\frac{l_c}{2 \,\textrm{kpc}}}
$$
The heavier  (but still light) nuclei bounded from Virgo might also be Li and Be:
$$
\delta_{rm-Be}\simeq{32^\circ}\left(\frac{Z}{Z_{\textrm{Be}^4}}\right)\left(\frac{6\cdot10^{19}\,\textrm{eV}}{E_{CR}}\right)\left(\frac{B}{3\,\mu\textrm{G}}\right)\sqrt{\frac{L}{20\, \textrm{kpc}}}\sqrt{\frac{l_c}{2\,\textrm{kpc}}}
$$

\begin{figure}[t]
\centering
\includegraphics[width=0.48\textwidth,clip]{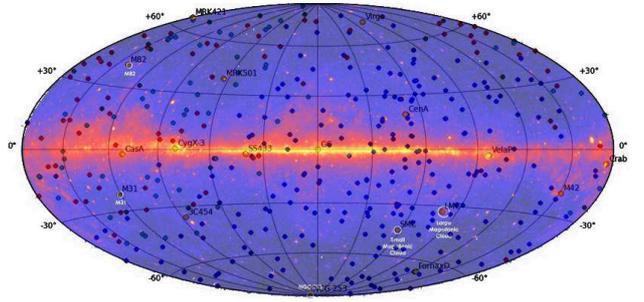}
\caption{Mollweide Projection in Celestial coordinates of all oldest and latest UHECR events from AGASA, AUGER and TA with several candidate sources labeled for a total of 376 UHECR events, over the Fermi map of MeV-GeV energies. There are clear signals around Vela and Cygnus and other galactic sources, as well as around Cen-A, M82 and Magellanic Clouds sources.}
\label{fig-7}
\end{figure}

It should be noted that the present anisotropy above the GZK energy of $5.5\cdot10^{19}$ eV (if extra-galactic)  \cite{Greisen:1966jv}   might leave a tail of signals: indeed the photo disruption of He into deuterium, tritium, He$^3$ and protons (and unstable neutrons), arising as clustered events at a half or a quarter (for the last most stable proton fragment) of the energy: \emph{protons having a quarter of the energy but a half of the charge of the He parent may form a tail  smeared around Cen-A at a twice larger angle} (\cite{Fargion2011nima}, \cite{Fargion2011c} and  \cite{Fargion2010}).
We suggested looking for correlated tails of events, possibly in  strings at relatively low energy $\sim (1.5 \div 3) \cdot10^{19}$ eV along the Cen-A train of events.
It should be noted that deuterium fragments have one half of the energy and mass of helium: \emph{therefore D and He spot are bent in the same way and overlap into UHECR circle clusters} \cite{Fargion2011c}.  Deuterium is  even more bounded in a very local Universe because of its fragility (explaining the absence of  Virgo). In conclusion, He like UHECRs may be bent by a characteristic angle as large as $\delta_{rm-He}\simeq16^\circ$; its expected lower energy deuterium or proton fragments at half energy ($30-25$ EeV) are also deflected accordingly at $\delta_{rm-p}\simeq16^\circ$; the last traces of protons at a quarter of the UHECR energy, around 20 EeV energy, will be bent and spread within $\delta_{rm-p}\simeq32^\circ$, exactly within the observed Cen-A UHECR multiplet \cite{Auger-Nov11}.

\section{TeV Gamma Rays and UHECRs}

In recent  updated UHECR AUGER-TA maps (fig.\ref{fig-2}, \ref{fig-3}, \ref{fig-4}, \ref{fig-5}) we have noted the first hint of a galactic source arising as a UHECR triplet \cite{Fargion2010}.
The hint of the Al$^{26}$ gamma map traced by Comptel somehow overlapping with UHECR events at 1-3 MeV, favors a role of UHECR radioactive elements (Al$^{26}$). The most prompt radioactive nuclei are  Ni$^{56}$, Ni$^{57}$, Co$^{56}$ and Co$^{60}$, produced by Supernova (SN) and possibly by their collimated GRB micro-jet components, ejecta in our own galaxy. Similar radioactive traces may arise by UHECR scattering on dense gas clouds.  Indeed in all SN Ia models, the decay chain  Ni$^{56}\rightarrow$ Co$^{56}\rightarrow $ Fe$^{56}$ provides the primary source of energy that powers the supernova optical display  days and even weeks following the explosion. Ni$^{56}$ decays by electron capture and the daughter Co$^{56}$  emits gamma rays by the nuclear de-excitation process; the two characteristic gamma lines are at $E_{\gamma}=158$ keV and $E_{\gamma}=812$ keV  respectively.

\begin{figure}[t]
\centering
\includegraphics[width=0.47\textwidth,clip]{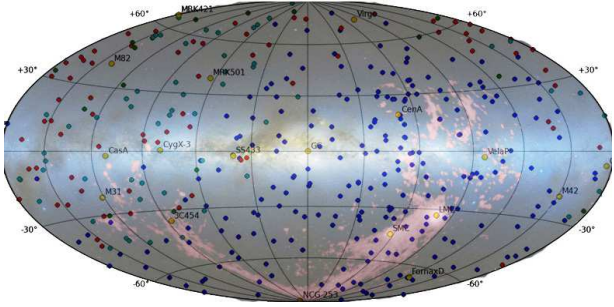}
\caption{All oldest and latest UHECR events from AGASA, AUGER and TA, in Galactic coordinates (Hammer Projection) with several candidate sources with label for a total 376 UHECR events; the map is overlapped on the Magellanic Stream in the nearby galactic volume.}
\label{fig-8}
\end{figure}
\begin{figure}[t]
\centering
\includegraphics[width=0.48\textwidth,clip]{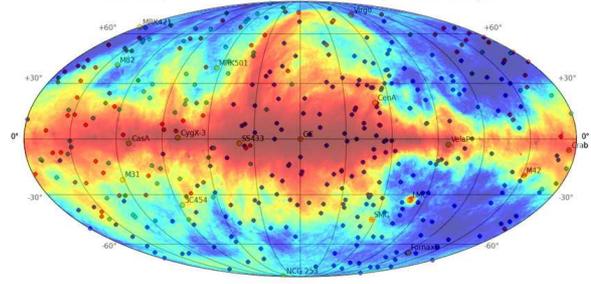}
\caption{All oldest  and latest UHECR events from AGASA, AUGER, TA, in galactic Malloweide coordinate with several candidate sources with label for a total 376 UHECR events, overlap on the galactic radiation at 408 MHz: it is remarkable that most of the events are located where there is synchrotron radio emission, possibly reinforcing a large galactic UHECR origin or a key role of UHECR shining into 408 MHz radio sky}
\label{fig-9}
\end{figure}
\begin{figure}[t]
\centering
\includegraphics[width=0.44\textwidth]{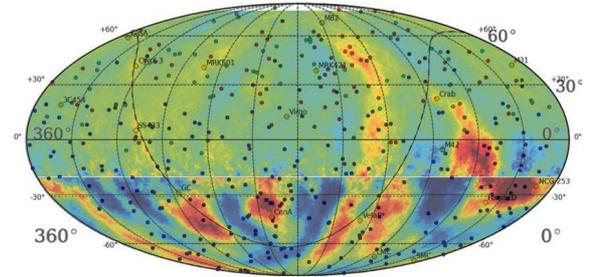}
\caption{All oldest  and latest UHECR events by AGASA, AUGER, TA, in galactic Hammer coordinate with several candidate sources with label.  We included 390 events = 231 AUGER UHECR, {72+15=87} TA, 58 AGASA and 14 (11 Haverah Park, 1 Yakutsk, 1 Volcano Ranch, 1 Fly's Eye - represented with black dots) over the gamma TeV background (2 TeV on the North, 20 TeV on the South), signals found in the Northern sky by ARGO in Tibet and in the Southern by ICECUBE at the south pole. Note that the ARGO detector is recording both gamma and Cosmic Ray at once, keeping partial memory of a small galactic plane anisotropy toward Cygnus X3.}
\label{fig-10}
\end{figure}

Their half lifetimes are spread from $35.6$ hours for Ni$^{57}$ to $6.07$ days for Ni$^{56}$. However, there are also more unstable radioactive rates, as  happens  for Ni$^{55}$ nuclei whose half life is just $0.212$s or for Ni$^{67}$, $21$s. Therefore we may have an apparent UHECR, boosted by a factor $\Gamma_{\textrm{Ni}^{56}}\simeq10^{9}$, lifetime spread from $2.12\cdot 10^{8}$ s or $6.7$ years (for Ni$^{55}$) up to nearly  $670$ years (for Ni$^{67}$), or even $4$ million years for Ni$^{57}$.
EeV and PeV radioactive UHECRs or their fragment may also play a role in gamma and neutrino emission (see UHECR events as in fig.\ref{fig-5}, \ref{fig-7}, \ref{fig-8} with different Fermi, IR, radio maps, fig.\ref{fig-9}-\ref{fig-10} for gamma TeV-UHECR correlation and in particular see fig.\ref{fig-9}, \ref{fig-10}, \ref{fig-magnetic2} in the TeV gamma background).
This consequent wide range of lifetimes guarantees a long life activity on  UHECR radioactive traces. The arrival tracks of these UHECR radioactive heavy nuclei may be  widely bent by galactic magnetic fields  as shown below.
Co$_{m}^{60}$  whose half life is $10.1$ min and whose decay gamma line is at $59$ keV is among the excited nuclei to mention  for the UHECR-TeV connection. For a boosted nominal Lorentz factor $\Gamma_{\textrm{Co}^{60}}=10^{9}$, we obtain $E_{\gamma}\simeq59$ TeV; note that a gamma air-shower exhibits a smaller secondary muon abundance with respect to the equivalent hadronic abundance; therefore a gamma  simulates ($10\%$)  a hadronic shower ($E_{gamma-hadron}\simeq6$~TeV)  corresponding closely to the observed ICECUBE-ARGO anisotropy \cite{ARGO}.
The decay boosted lifetime is $19000$ years, corresponding to  6 kpc distance. Therefore Co$_{m}^{60}$ energy decay traces, lifetime and spectra fit well within  the present UHECR-TeV connection for nearby galactic sources such as Vela and - probably - the Crab. Other radioactive scattering traces, usually at lower energy may also shine at hundreds or tens of TeV or below by Inverse Compton and synchrotron radiation. Therefore their UHECR bent parental nuclei may  also shine in TeV Cosmic Ray signals. Electrons and neutrinos are also born in $\beta$-decay processes providing a new diffused gamma and PeV neutrino source. Also light nuclei such as He$^{6}$ might decay in flight playing a radioactive UHECR-TeV role.

\begin{figure}[t]
\centering
\includegraphics[width=0.43\textwidth]{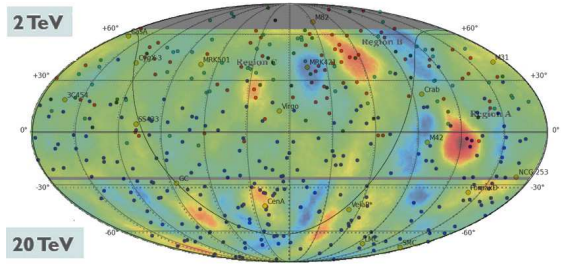}
\caption{Same as the above Fig. \ref{fig-10}. Note that Hawk, like older Milagro  detector, is recording  mostly Cosmic Ray (not gamma); therefore they do not see (clearly like ARGO, figure above) the galactic anisotropy toward Cygnus X3.}
\label{fig-11}
\end{figure}

\section{ UHECR galactic bending  for Ni$^{57}$}

Cosmic Rays are blurred by magnetic fields. UHECRs also suffer from Lorentz force deviations. This smearing could be a source of UHECR features, mostly along the Cen-A direction. There are at least three mechanisms for magnetic deflection along the galactic plane, giving rise to a sort of galactic spectroscopy of UHECRs \cite{Fargion2008}.
Magnetic  bending by extra-galactic fields is in general negligible in comparison with the galactic field.  Late nearby (almost local) bending by a near coherent galactic arm field, random bending by turbulence and random deflection along the whole plane inside different arms are:
\begin{enumerate}[leftmargin=0cm,itemindent=.5cm,labelwidth=\itemindent,labelsep=0cm,align=left]
\item The coherent Lorentz  bending angle, $\delta_{Coh}$, of a proton UHECR (or nucleus)  (above the GZK energy) within a galactic magnetic field in a final nearby coherent length of $l_c = 1$ kpc is
$$
\delta_{Coh-p}\simeq{2.3^\circ}\cdot \frac{Z}{Z_{H}}\left(\frac{6\cdot10^{19}\,\textrm{eV}}{E_{CR}}\right)\left(\frac{B}{3\,\mu\textrm{G}}\right)\left(\frac{l_c}{\textrm{kpc}}\right).
$$
\item Random bending by random turbulent magnetic fields, whose coherent sizes (tens of parsecs) are short and whose final deflection angle is  smaller than others, are ignored here.
\item The ordered multiple UHECR deflection along the galactic plane across and by alternate arm magnetic field directions whose final random deflection angle is remarkable and discussed below.
The bending angle value is  quite different for a heavy nucleus such as a UHECR from Vela whose distance is only $0.29$ kpc:
 $$
 \delta_{Coh-\textrm{Ni}}\simeq18.7^\circ\cdot\frac{Z}{Z_{\textrm{Ni}^{28}}}\left(\frac{6\cdot10^{19}\,\textrm{eV}}{E_{CR}}\right)\left(\frac{B}{3\,\mu \textrm{G}}\right)\left(\frac{l_c}{0.29\,\textrm{kpc}}\right)
 $$
 \end{enumerate}

\begin{figure}[t]
\centering
\includegraphics[width=0.47\textwidth,clip]{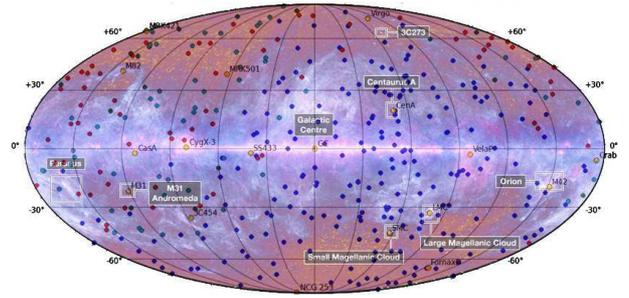}
\caption{All oldest and latest UHECR events by AGASA, AUGER, TA, in galactic Hammer coordinates with several candidate sources with label for a total of 376 UHECR events, superposed to the Planck map of  infrared due to dust. Again there is a remarkable signal in the absence of dust where there are also none or few  UHECR events.}
\label{fig-12}
\end{figure}

\begin{figure}[t]
\centering
\includegraphics[width=0.48\textwidth,clip]{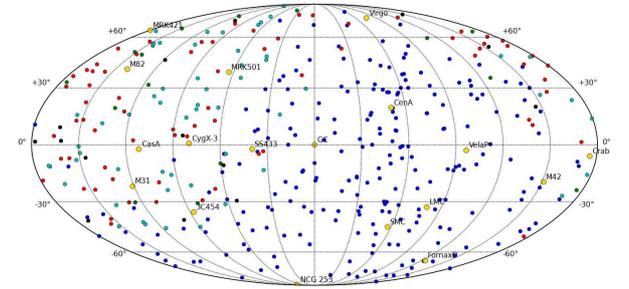}
\caption{As above, in Mollweide Projection, Galactic Coordinate map for 231 AUGER UHECR, {72+15=87} TA, 58 AGASA and 14 (11 Haverah Park, 1 Yakutsk, 1 Volcano Ranch, 1 Fly's Eye - represented with black dots) additional records \cite{Nagano} \cite{Ave}. Few potential sources are labeled.}
\label{fig-13}
\end{figure}

  Note that this spread can explain the nearby Vela TeV anisotropy area (because of the in-flight radioactive emission) around its correlated UHECR triplet.  There is a further extreme possibility: a Crab-like pulsar at a few kpc is feeding the TeV anisotropy, as shown by overlap clustering of UHECR events, enclosed in a black oval dashed perimeter in Figs. \ref{fig-magnetic} and \ref{fig-magnetic2}. For  Crab distances the galactic bending is:
  $$
  \delta_{Coh-\textrm{Ni}}^\prime\simeq129^\circ\cdot\frac{Z}{Z_{\textrm{Ni}^{28}}}\left(\frac{6\cdot10^{19}\,\textrm{eV}}{E_{CR}}\right)\left(\frac{B}{3\,\mu \textrm{G}}\right)\left(\frac{l_c}{2\,\textrm{kpc}}\right)
  $$
 Such a spread is, again, able to explain the localized TeV anisotropy produced in the Crab (2 kpc), apparently  extending  around an area near Orion, where  the spread UHECR events also seem to be clustered. Such heavy iron-like  (Ni, Co) UHECRs  are mostly bounded inside a Galaxy, as well as in a Virgo cluster  because of the big charge and large angle bending, possibly explaining the  absence of UHECRs in that direction. The possible galactic component of UHECR is suggested by the correlated dark Hydrogen and dust map with the UHECR distribution as well as radio 408 MHz emission:  see Figures \ref{fig-9}-\ref{fig-12}.

\begin{figure}[t]
\centering
\includegraphics[width=0.48\textwidth,clip]{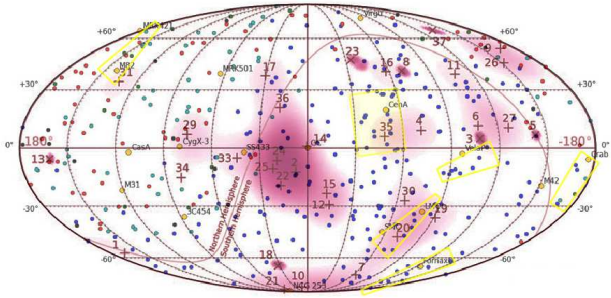}
\caption{All oldest UHECR by AGASA, AUGER, TA, in galactic Mollweide coordinate with several candidate sources with label: blue AUGER, red 72 events by TA, green last 15 by TA,  58 old AGASA cyan events for a total of 376 UHECR events. Note the crowding of triplet around the M82 source, the multiplet around Cygnus X3, M82, ss433, NGC253. The possible correlation between UHECR and ICECUBE UHE neutrino is still questionable. We note anyway a doublet near Vela and other marginal correlation discussed elsewhere.}
\label{fig-15}
\end{figure}

\begin{figure}[t]
	\centering
	\includegraphics[width=0.47\textwidth]{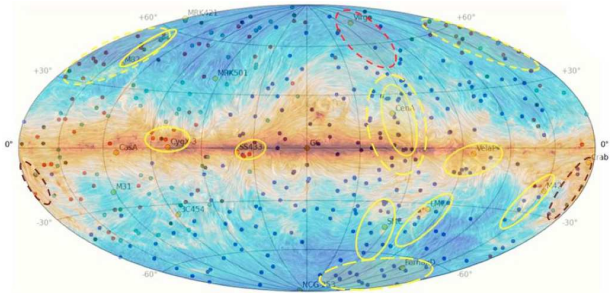}
	\caption{All UHECR events superimposed to the polarized emission in Planck data by Milky Way. Note the oval contour that defines possible UHECR clustering along candidate source. Photo Credit: ESA. We added to the 376 events 14 events by Haverah Park, Fly's Eye, Yakutsk. Other events by HiRes has not been included. The map is in galactic Hammer coordinates.}
	\label{fig-magnetic}
\end{figure}

\section{Conclusions}
\label{sec-4}
It is worth remembering that in an (unexpected)  future where UHECRs are more and more homogenous and uncorrelated with nearby GZK sources they will require a far cosmic UHECR origin (as  has been the case for GRB).
In this (still not actual) frame the most reasonable candidates for UHECR
are  UHE (ZeV) neutrinos  (considered nearly 20 years ago)
from far away cosmic source edges (AGN, GRBs jets), evading the GZK cutoff; they collide with
relic neutrinos (with mass) clustered around our Galaxy's halo. These scatterings
lead to the Z resonance and its decay to nucleon secondaries observed as UHECRs \cite{Fargion1997}.
A relic neutrino mass of 0.4 eV may be ideal for such a resonance,
even if a 0.2-0.1 eV neutrino mass may still be (better) compatible with Planck limits
and comparable with atmospheric mass splitting \cite{Far}.
A fourth (now in fashion, but still speculative) eV sterile neutrino, being thermalized in the big bang and better clustered,  fits  better the Z-showering solution. The associated tau neutrino component (either by Z decay or GZK cut off)
 and its possible (yet unobserved) double bang \cite{Learned}, in ICECUBE
or any  consequent PeV-EeV Tau airshowers in TA and AUGER maybe the road map to test any
UHECR and the UHE neutrino astronomy connections (see \cite{FarTau}, \cite{Fargion006},  \cite{Fargion1999}, \cite{Auger-01}, \cite{Feng02}, \cite{Auger07}, \cite{Auger08}).

\begin{figure}[t]
	\centering
	\includegraphics[width=0.47\textwidth]{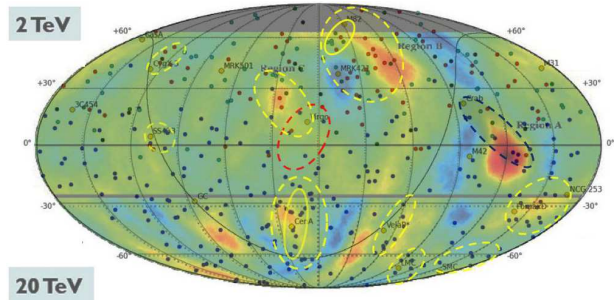}
	\caption{As in fig.\ref{fig-magnetic}, in celestial coordinates (Mollweide), over TeVs gamma anisotropy by Hawk and IceCube detectors. Note in particular the possible asymmetric clustering on UHECR nearby Virgo as well as other oval contour for additional candidates. We added to the 376 events 14 events by Haverah Park, Fly's Eye, Yakutsk. Other events by HiRes have  not been included.}
	\label{fig-magnetic2}
\end{figure}

 In conclusion, the two main clusterings in the NORTHERN (HOT SPOT by TA) and in the SOUTHERN
 (HOT SPOT by AUGER) are too wide (15-20 degree, as seen in fig.\ref{fig-magnetic}, \ref{fig-magnetic2}) to be due to protons
(whose bending extend just a few degree)  as proposed  by TA.
Such Hot Spots are at the same time too narrow to be produced  by  cosmic
UHECR heavy (Fe, Ni) nuclei (whose bending may exceed 80 degrees), as proposed by AUGER. Light or the lightest nuclei may be the best compromise as the natural extragalactic UHECR carrier (\cite{Fargion2011nima}, \cite{Fargion2008},  \cite{Fargion2011c}, \cite{Fargion2010}, \cite{Fargion09a}).
They fit or coexist with AUGER and TA composition results. Also a few, very nearby galactic  UHECR heavy-light
 nuclei may spread over a few tens of degree clustering spots.

We have tried  here to present updated maps to correlate these
first few clusterings with their
source, offering  a name or a preliminary identity. We combined all known UHECR archives
(AUGER-TA-AGASA and a few more UHECR detections)  in best known usable coordinates (see fig. \ref{fig-2}-\ref{fig-5}, \ref{fig-12}).
We overlap UHECR on the Fermi gamma sky, on Radio or IR maps, (see fig. \ref{fig-7}-\ref{fig-11})
as well as with recent  Planck magnetic maps and ARGO-ICECUBE TeV maps (see fig \ref{fig-magnetic}-\ref{fig-magnetic2})
tracking the oval clustering mask areas.
Analogous recent attempts to correlate UHECR within any wider (canonical)
 GZK volumes (with AGN, BL Lac sources) failed \cite{Auger-Nov14}.

Therefore the best  clustered source event is near Cen-A, in the SOUTHERN HOT SPOT; but also
M82   might feed the observed TA clustering
 by light nuclei whose bending  shines the NORTHERN HOT SPOT \cite{Fargion2014}.
The incoherent random bending in Cen-A, may set the source in the main
clustering center; because of the horizontal spiral arms fields the spread occurs in a vertical (orthogonal to the galactic plane) direction. Otherwise a coherent bending in the M82 case may lead the  source at the edge of the
clustering Hot Spot; this occurs because the unique magnetic field coherently bends  positive charges in the same asymmetric way by Lorentz forces. The same coherent asymmetric bending could be responsable
for the Crab location at the oval edge, Fornax and  NGC253 clustering, the Vela asymmetric
position respect the nearby 8 events (see  fig.\ref{fig-magnetic}-\ref{fig-magnetic2}).
The lightest nuclei provide at the moment a natural  bending angle (and a safe cut off for UHECR)  from Virgo,
whose distance is too far and whose central disk sky is nearly empty (see  fig \ref{fig-magnetic2}, in celestial coordinates).
However, as in  fig.\ref{fig-magnetic2}, let us note a possible remarkable   weak
asymmetric clustering linking bent UHECR events
to Virgo in a nearby area (see  fig.\ref{fig-magnetic2}).
The probability that such
a clustering of 13 events in 159 takes place  in the Northern
sky in a thousand square degree  oval  is less than  $7\cdot10^{-3}$.
This area is coincident with the C anisotropy area found by ARGO (fig.\ref{fig-magnetic2}).

The probability to observe 8 events within the nearby
Vela area (fig.\ref{fig-magnetic}) is only $2.5\cdot10^{-2}$;
however, the corresponding clustering probability around the thin Crab oval area of 11 events
is near $10^{-4}$ (fig. \ref{fig-magnetic2}). The probability to find 7 events around Cygnus X3
is $3.1\cdot10^{-3}$; note that two additional HIRES events are in this Cygnus area
(but their exact coordinates have not been  published) leading to a remarkable correlation,
as well as the ARGO anisotropy around that area.
The probability to find in a very narrow oval area four UHECR
(two of them being the most energetic ones of TA and AUGER),
pointing to SS433 (or AqX1) is about  $6.7\cdot10^{-3}$, see \cite{Fargion2014}, \cite{Fargion13}.
The eventual clustering around the Small and Large Magellanic Clouds  (SMC.LMC) is
still a weak but interesting hint being as rare as  $3.5\cdot10^{-2}$.

In conclusion the two nearest AGNs, Cen-A and M82, may explain
the main strong statistical presence of two Hot Spots \cite{Fargion2014}.
The probability for the by-chance event for Cen-A (wide dashed Hot Spot)
is around ($1.5\cdot10^{-5}$), the probability for its inner smaller clustering
is about $5.4\cdot10^{-4}$, (fig. \ref{fig-magnetic}-\ref{fig-magnetic2}). We remember also  the
tens EeV clustering along the same oval area (fig. \ref{fig-1}).
The probability to have a Northern Hot Spot (narrow area around M82)
with 8 events is $2.3\cdot10^{-3}$ while the probability for
the whole wide Northern Hot Spot (dashed lines in fig.\ref{fig-magnetic}-\ref{fig-magnetic2}) is $1.3\cdot10^{-4}$ \cite{Fargion2014}, \cite{FarO}.
The lightest UHECR nuclei (He, Li, Be, B)  may be  the main natural carriers.
Their fragments ($\alpha$, $\gamma$), by their decay, as being radioactive (or by photo dissociation in flight via infra-red radiations) may be the cause of the correlated TeV anisotropy. A few heavier (Ni, Co) UHECR nuclei by nearby galactic sources as  Vela, Cygnus and Crab, Large Magellanic and Small Magellanic Clouds
 may also play a role.

Additional samples of UHECR events (as well as multiplets at tens EeV energies), or more precise TeV anisotropy maps, might confirm (or reject) our present tentative model based on the lightest nuclei partially coherent or randomly deflected and mostly originating in a nearby Universe.

Indeed, for instance, the very recent results by PAO with more data \cite{PAO}
is correlated in most UHECR events composition (see their fig.3) with light or lightest nuclei (and not with p, Fe ones); moreover  above 8 EeV energy  AUGER discovered a remarkable dipolar anisotropy (in the wider Southern Hot Spot) at 4 sigma (see their Fig.6):
this  UHECR spot  encompasses or well correlates with the higher energy one in our tagged area (see  Fig \ref{fig-magnetic}), along Cen A, as we expected, within a nearby Universe.

\end{document}